\documentclass{article}%
 
\usepackage{graphicx}
\usepackage{color}
\usepackage{tikz}
\usepackage{amssymb}
\usepackage{caption2} 
\usepackage{lmodern}
\usepackage[T1]{fontenc}
\usepackage{hyperref}
\usepackage{algorithm,algpseudocode}

\newtheorem{theorem}{Theorem}
\newtheorem{lemma}{Lemma}
\newtheorem{definition}{Definition}
\newtheorem{operation}{Operation}
\newenvironment{proof}{{\sc Proof. }}{\hfill$\Box$\vspace{0.1in}}
\newcommand{\mc}[1]{{\cal {#1}}}

\usepackage{geometry} 
\title{An improved local search based algorithm for $k^-$-star partition}
\author{Mingyang Gong\footnote{Gianforte School of Computing, Montana State University, Bozeman, MT 59717, USA.
Email: {\tt \{mingyang.gong, brendan.mumey\}@montana.edu}}
\and
Guohui Lin\footnote{Department of Computing Science, University of Alberta, Edmonton, Alberta T6G 2E8, Canada.
Email: {\tt guohui@ualberta.ca}}
\and 
Brendan Mumey$^*$
}

\begin{document}
\maketitle
\begin{abstract}
We study the $k^-$-star partition problem that aims to find a minimum collection of vertex-disjoint stars, each having at most $k$ vertices to cover all vertices in a simple undirected graph $G = (V, E)$.
Our main contribution is an improved $O(|V|^3)$-time $(\frac k2 - \frac {k-2}{8k-14})$-approximation algorithm.

Our algorithm starts with a $k^-$-star partition with the least $1$-stars and a key idea is to distinguish critical vertices, 
each of which is either in a $2$-star or is the center of a $3$-star in the current solution.
Our algorithm iteratively updates the solution by three local search operations 
so that the vertices in each star in the final solution produced cannot be adjacent to too many critical vertices.
We present an amortization scheme to prove the approximation ratio in which the critical vertices are allowed to receive more tokens from the optimal solution.

\end{abstract}




\section{Introduction}\label{sec:intro}

Given a simple and undirected graph $G = (V, E)$, a star in $G$ is a single vertex, an edge or 
a connected subgraph of $G$ such that exactly one vertex is of degree at least $2$ and the other vertices are of degree~$1$.
An $\ell$-star ($\ell^+$-star, $\ell^-$-star, respectively) is a star with exactly (at least, at most, respectively) $\ell$ vertices.
The center of a star is the vertex with the maximum degree and except the center, the other vertices are called satellites.
For a $2$-star, i.e., an edge, we arbitrarily choose one vertex as its center and the other one as the satellite.
Therefore, each star has exactly one center.
A path in $G$ is a subgraph of $G$ such that the vertices can be ordered into a sequence $v_1, \ldots, v_\ell$ and the edge set is $\{ \{ v_j, v_{j+1} \}: j =1, \ldots, \ell-1 \}$.
Similarly to an $\ell$-, $\ell^+$ and $\ell^-$-star, we can define an $\ell$-, $\ell^+$ and $\ell^-$-paths.
For a given positive integer $k$, the $k^-$-star ($k^-$-path, respectively) partition problem seeks a minimum collection of vertex-disjoint $k^-$-stars ($k^-$-paths) to cover all the vertices in $G$,
which is denoted as $k$SP ($k$PP, respectively) for short.
Note that an $\ell$-star with $\ell \in \{ 1, 2, 3 \}$ is the same as an $\ell$-path.
Therefore, $k$SP is the same as $k$PP when $k \in \{ 1, 2, 3 \}$.

We next review the related works of $k$PP.
Clearly, $1$PP is trivial and when $k = 2$, the problem is equivalent to finding a maximum matching, 
which can be done in $O(\sqrt{|V|} |E|)$ time~\cite{MV80}.
But when $k \ge 3$, $k$PP becomes NP-hard~\cite{GJ79}.
For $3$PP, Monnot and Toulouse presented a matching-based algorithm that achieves $\frac 32$-approximation~\cite{MT07}.
Chen et al. observed that minimizing the number of $1$-paths in a $3$PP is polynomial-time solvable~\cite{CGL19a}
and by further merging three $2$-paths into two $3$-paths in the computed solution, 
they developed a $\frac {13}9$-approximation algorithm,
which was improved to $\frac 43$~\cite{CGL22} by considering more $2$- and $3$-paths.
Finally, Chen et al.~\cite{CGS19} designed a novel weight schema which carefully assigns a weight to each path in the current solution and 
presented the current best $\frac {21}{16}$-approximation algorithm.
For $k$PP with a general $k$, Chen et al.~\cite{CGL19a} presented a $\frac k2$-approximation algorithm by minimizing the $1$-paths in the collection.
Afterwards, Li et al.~\cite{LYL24a} presented a local search based algorithm, which repeatedly reduces the number of paths in the current solution or; generates a $(k-1)$-path or a $k$-path
and brought down the ratio to $\frac {k^2}{3(k-1)} + \frac {k+3}{6(k-1)}$.
The state-of-the-art result is from~\cite{LYL24b} where the authors employ the $\frac 67$-approximation algorithm for the {\em maximum path cover} problem~\cite{BK06}
and; cut each $(k+1)^+$-path into a set of $k$-paths and a possible $(k-1)^-$-path.

Recall that $k$PP is targeted at short paths and its complementary version is to find $k^+$-paths, i.e., long paths. 
One sees that it may be impossible to cover all the vertices by long paths and thus Gong et al.~\cite{GEF24} switched to consider another problem, denoted as MPC$^{k+}$ for a fixed positive integer $k$,
which aims to cover most vertices by a collection of vertex-disjoint $k^+$-paths.
Clearly MPC$^{1+}$ is trivial.
MPC$^{2+}$ equates to finding a path partition with the least $1$-paths and is also polynomial-time solvable~\cite{CGL19a}.
Note that each vertex of a $3^+$-path is of degree at most $2$ and there exists at least one vertex with degree exactly $2$,
which is exactly the definition of a {\em $2$-piece}~\cite{HHS06}.
In fact, MPC$^{3+}$ can be solved in polynomial time by the algorithm presented in~\cite{HHS06} for the maximum $2$-piece problem.
Therefore, when $k \in \{ 1, 2, 3 \}$, the MPC$^{k+}$ problem is in P.
But when $k \ge 4$, it becomes NP-hard~\cite{KLM23}.
Kobayashi et al. presented a $4$-approximation algorithm for MPC$^{4+}$, which was gradually improved to $2$~\cite{GEF24}, then $1.874$~\cite{GCL23} and eventually $\frac 53$~\cite{GCL25}.
The state-of-the-art result is $2.511$~\cite{GCL24} for MPC$^{5+}$ and $0.4394k + O(1)$~\cite{GEF24} for a general $k \ge 6$, respectively.

Recall that $k$PP is the same as $k$SP if $k \in \{ 1, 2, 3 \}$ and thus we assume $k \ge 4$ and review the works related to $k$SP.
For $4$SP, Bao et al. presented a $1.9$-approximation algorithm~\cite{BYL24} and for a general $k$, there is a $\frac k2$-approximation algorithm~\cite{HK86,BG11} by minimizing the number of $1$-stars in the found collection.
Recently, Xu et al.~\cite{XYL25} presented an improved $O(|V|^5 + |V|^{k/2})$-time $(\frac k2 - \frac {k-2}{k(k+1)})$-approximation algorithm when $k \ge 6$ is even
and an $O(|V|^5)$-time $(\frac k2 - \frac {k-2}{2k^2})$-approximation algorithm when $k \ge 5$ is odd, respectively.
The algorithm computes a $k^-$-star partition with the least $1$-stars and update the solution by two or three local improvement operations, each of which reduces the number of stars by at least one,
until none of them is applicable.
In~\cite{HZC24}, the authors studied a star packing problem that seeks to cover most vertices by a vertex-disjoint $\ell$-stars where $\ell \le k$ but $\ell \ne t$ for two given integers $3 \le t \le k$.
They presented an approximation algorithm by local search, which achieved $\frac 1{t+1+1/k}$-approximation.
Similarly to MPC$^{k+}$, the authors in~\cite{HZC24} also considered another star packing problem 
to cover most vertices by $k^+$-stars 
and presented several algorithms for a general $k$ or a specific $k$.


\subsection{Our results}

In this paper, we study the $k$SP problem and our main contribution is an $O(|V|^3)$-time 
$(\frac k2 - \frac {k-2}{8k-14})$-approximation algorithm for every $k \ge 4$,
which is strictly better than the $(\frac k2 - \frac {k-2}{k(k+1)})$- and $(\frac k2 - \frac {k-2}{2k^2})$-approximation algorithm in~\cite{XYL25}.
Our algorithm is local search based and the novel idea is that we define each vertex in a $2$-star or the center of a $3$-star as a critical vertex.
Generally speaking, our algorithm uses three local search operations to gradually reduce the number of critical vertices
(rather than the number of stars in the solution as in~\cite{XYL25})
so that the vertices of a star in the computed solution cannot be adjacent to too many critical vertices.
Note that our algorithm runs in $O(|V|^3)$ time for every $k \ge 4$ and thus is much faster than the ones in~\cite{XYL25} especially for even $k$'s.
For the approximation ratio, we present an amortization scheme in which each vertex can receive certain number of tokens from the optimal solution but critical vertices are allowed to receive more.
Therefore, all the cases can be well-balanced and we finally show that the approximation ratio is bounded by 
$\frac k2 - \frac {k-2}{8k-14}$.

\section{Preliminaries}
\label{sec2}

We fix a simple undireced graph $G = (V, E)$ for discussion where $V$ is the vertex set and $E$ is the edge set, respectively.
Let $n = |V|$ and $m = |E|$.
An edge in $E$ connects two vertices $u, v \in V$ and is represented as $\{ u, v \}$.
Given a graph $G$, we also use $V(G)$ and $E(G)$ to denote the vertex set and edge set of $G$, respectively.
A graph $G'$ is a {\em subgraph} of $G$ if $V(G') \subseteq V$ and $E(G') \subseteq E$.
The number of vertices in $G'$ is defined as the {\em order} of $G'$.
The degree of a vertex $v \in V(G)$ is the number of edges in $G$ incident to $v$.

We denote by $S = v_1$-$v_2 \ldots v_\ell$ an {\em $\ell$-star} where the order of $S$ is exactly $\ell$
and there is an edge connecting $v_1$ and $v_j$ for every $j = 2, \ldots, \ell$.
The vertex $v_1$ is defined as the {\em center} of $S$ and the other vertices $v_2, \ldots v_\ell$ are {\em satellites}.
One sees that a $1$-star has no satellite and for any $2$-star, we can arbitrarily choose one vertex as the center and the other one as the satellite.
This way, we can assume that a star has exactly one center.
Similarly to an $\ell$-star, we say a star is an $\ell^-$-star ($\ell^+$-star, respectively) if its order is at most (at least, respectively) $\ell$.

Given a positive integer $k$, a {\em $k^-$-star partition} $\mc{S}$ of $G$ is a collection of vertex-disjoint $k^-$-stars 
such that $V(\mc{S}) = V(G)$.
Our problem aims to find a $k^-$-star partition such that the number of stars in the collection is minimized.
In the following, we assume that $\mc{S}$ is the $k^-$-star partition produced by a specific algorithm
and we let $\mc{S}_j \subseteq \mc{S}$ denote the subset of $j$-stars in $\mc{S}$ for every $j = 1, \ldots, k$.
Similarly, we fix an optimal $k^-$-star partition $\mc{Q}$ for discussion and let $\mc{Q}_j \subseteq \mc{Q}$ denote the subset of $j$-stars in $\mc{Q}$ for every $j = 1, \ldots, k$.
For ease of presentation, we say a vertex is a center (satellite, respectively) of $\mc{S}$ ($\mc{Q}$, respectively) if it is the center (a satellite, respectively) of some star in $\mc{S}$ ($\mc{Q}$, respectively).

\section{The algorithm}
\label{sec3}

Our algorithm starts with a $k^-$-star partition $\mc{S}$ with the minimum number of $1$-stars~\cite{HK86,BG11},
which can be obtained in $O(nm)$ time.
Recall that the objective of $k^-$-star partition is to minimize the number of stars in the collection $\mc{S}$
and the initial $\mc{S}$ minimizes the number of $1$-stars.
Before we present the operations, we give the following definition for critical vertices.

\begin{definition}
\label{def01}
(Tiny stars and critical vertices)
Given the current $k^-$-star partition $\mc{S}$, let $\mc{T} = \mc{S}_2 \cup \mc{S}_3$
and each star in $\mc{T}$ is defined as {\em tiny}.

A vertex is {\em critical} if it appears in $V(\mc{T})$ but is not a satellite of a $3$-star.
We say two critical vertices are {\em separate} if they appear in different stars in $\mc{S}$.
\end{definition}

By Definition~\ref{def01}, a critical vertex is either in a $2$-star or is the center of a $3$-star.
Moreover, the two vertices in a $2$-star of $\mc{S}$ are not considered as separate.

\begin{definition}
\label{def02}
We define a quantity $q(\mc{S}) = 3|\mc{S}_2| + |\mc{S}_3|$, which indicates the weighted number of critical vertices in $\mc{S}$.
\end{definition}

In the following, we present three local operations for a general $k$ to minimize the quantity $q(\mc{S})$
but no new $1$-stars are generated.

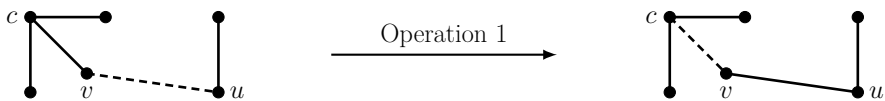
\begin{figure}[thb]
\begin{center}
\begin{tikzpicture}[scale=0.5,transform shape]

\draw [thick, line width = 1pt] (-8, -10) -- (-6, -10);
\draw [thick, line width = 1pt] (-8, -10) -- (-8, -12);
\draw [thick, line width = 1pt] (-8, -10) -- (-6.5, -11.5);
\filldraw (-8, -10) circle(.15);
\filldraw (-6, -10) circle(.15);
\filldraw (-8, -12) circle(.15);
\filldraw (-6.5, -11.5) circle(.15);
\node[font=\fontsize{20}{6}\selectfont] at (-8.5, -10) {$c$};
\node[font=\fontsize{20}{6}\selectfont] at (-6.5, -12) {$v$};

\draw [densely dashed, line width = 1pt] (-6.5, -11.5) -- (-3, -12);

\draw [thick, line width = 1pt] (-3, -10) -- (-3, -12);
\filldraw (-3, -10) circle(.15);
\filldraw (-3,-12) circle(.15);
\node[font=\fontsize{20}{6}\selectfont] at (-2.5, -12) {$u$};

\draw [-latex, thick] (0, -11) to (6, -11);
\node[font=\fontsize{20}{6}\selectfont] at (3, -10.5) {Operation~1};

\draw [thick, line width = 1pt] (9, -10) -- (11, -10);
\draw [thick, line width = 1pt] (9, -10) -- (9, -12);
\draw [densely dashed, line width = 1pt] (9, -10) -- (10.5, -11.5);
\filldraw (9, -10) circle(.15);
\filldraw (11, -10) circle(.15);
\filldraw (9, -12) circle(.15);
\filldraw (10.5, -11.5) circle(.15);
\node[font=\fontsize{20}{6}\selectfont] at (8.5, -10) {$c$};
\node[font=\fontsize{20}{6}\selectfont] at (10.5, -12) {$v$};

\draw [thick, line width = 1pt] (10.5, -11.5) -- (14, -12);

\draw [thick, line width = 1pt] (14, -10) -- (14, -12);
\filldraw (14, -10) circle(.15);
\filldraw (14,-12) circle(.15);
\node[font=\fontsize{20}{6}\selectfont] at (14.5, -12) {$u$};

\end{tikzpicture}
\end{center}
\caption{An illustration of Operation~\ref{op01} where the dashed (thick, respectively) edges are in (not in, respectively) $\mc{S}$.
Operation~\ref{op01} revises a $2$-star and a $4^+$-star and; extracts a $3$-star and a $3^+$-star.
\label{fig01}}
\end{figure}

The first operation is applicable when there is an edge $\{ u, v \}$ such that $u$ appears in a $2$-star and $v$ is a satellite of a $4^+$-star centered at $c$ in $\mc{S}$.
Then the operation replaces the edge $\{ c, v \}$ by $\{ u, v \}$ so that a $3$-star and a $3^+$-star are formed.
See Figure~\ref{fig01} for an illustration.

\begin{operation}
\label{op01}
Given an edge $\{ u, v \}$ where $u$ is in a $2$-star and $v$ is a satellite of a $4^+$-star centered at $c$,
the operation replaces $\{ c, v \}$ by $\{ u, v \}$.
\end{operation}

\begin{lemma}
\label{lemma01}
Determining whether or not Operation~\ref{op01} is applicable can be done in $O(m)$ time
and if it is applicable, then $q(\mc{S})$ is reduced by at least $1$.
\end{lemma}
\begin{proof}
It is sufficient to determine whether Operation~\ref{op01} is applicable by scanning every edge in $G$.
Therefore, Operation~\ref{op01} can be done in $O(m)$ time.
Clearly, if Operation~\ref{op01} is executed, then $|\mc{S}_2|$ is reduced by exactly one and one sees that $|\mc{S}_3|$ may be increased by at most $2$ (in Figure~\ref{fig01}).
Therefore, $q(\mc{S})$ is decreased by at least one.
\end{proof}

\begin{figure}[thb]
\begin{center}
\begin{tikzpicture}[scale=0.38,transform shape]

\draw [thick, line width = 1pt] (-8, -10) -- (-8, -12);
\draw [thick, line width = 1pt] (-8, -10) -- (-6.5, -11.5);
\filldraw (-8, -10) circle(.15);
\filldraw (-8, -12) circle(.15);
\filldraw (-6.5, -11.5) circle(.15);
\node[font=\fontsize{20}{6}\selectfont] at (-8.5, -10) {$v_1$};
\node[font=\fontsize{20}{6}\selectfont] at (-6.4, -11) {$v_3$};
\node[font=\fontsize{20}{6}\selectfont] at (-7.5, -12) {$v_2$};

\draw [densely dashed, line width = 1pt] (-6.5, -11.5) -- (-3, -12);
\draw [densely dashed, line width = 1pt] (-8, -12) -- (-12, -10);
\draw [densely dashed, line width = 1pt] (-8, -10) -- (-8, -8);

\draw [thick, line width = 1pt] (-8, -8) -- (-6, -8);
\filldraw (-8, -8) circle(.15);
\filldraw (-6, -8) circle(.15);
\node[font=\fontsize{20}{6}\selectfont] at (-8.6, -8) {$w_1$};
\draw [thick, line width = 1pt] (-3, -10) -- (-3, -12);
\filldraw (-3, -10) circle(.15);
\filldraw (-3,-12) circle(.15);
\node[font=\fontsize{20}{6}\selectfont] at (-2.4, -12) {$w_3$};
\draw [thick, line width = 1pt] (-12, -10) -- (-12, -12);
\draw [thick, line width = 1pt] (-12, -10) -- (-11, -11.8);
\filldraw (-12, -10) circle(.15);
\filldraw (-12,-12) circle(.15);
\filldraw (-11, -11.8) circle(.15);
\node[font=\fontsize{20}{6}\selectfont] at (-12, -9.5) {$w_2$};


\draw [-latex, thick] (0, -10.5) to (6, -10.5);
\node[font=\fontsize{20}{6}\selectfont] at (3, -10) {Operation~2};


\draw [densely dashed, line width = 1pt] (13, -10) -- (13, -12);
\draw [densely dashed, line width = 1pt] (13, -10) -- (14.5, -11.5);
\filldraw (13, -10) circle(.15);
\filldraw (13, -12) circle(.15);
\filldraw (14.5, -11.5) circle(.15);
\node[font=\fontsize{20}{6}\selectfont] at (12.5, -10) {$v_1$};
\node[font=\fontsize{20}{6}\selectfont] at (14.6, -11) {$v_3$};
\node[font=\fontsize{20}{6}\selectfont] at (13.5, -12) {$v_2$};

\draw [thick, line width = 1pt] (14.5, -11.5) -- (18, -12);
\draw [thick, line width = 1pt] (13, -12) -- (9, -10);
\draw [thick, line width = 1pt] (13, -10) -- (13, -8);

\draw [thick, line width = 1pt] (13, -8) -- (15, -8);
\filldraw (13, -8) circle(.15);
\filldraw (15, -8) circle(.15);
\node[font=\fontsize{20}{6}\selectfont] at (12.4, -8) {$w_1$};
\draw [thick, line width = 1pt] (18, -10) -- (18, -12);
\filldraw (18, -10) circle(.15);
\filldraw (18,-12) circle(.15);
\node[font=\fontsize{20}{6}\selectfont] at (18.6, -12) {$w_3$};
\draw [thick, line width = 1pt] (9, -10) -- (9, -12);
\draw [thick, line width = 1pt] (9, -10) -- (10, -11.8);
\filldraw (9, -10) circle(.15);
\filldraw (9,-12) circle(.15);
\filldraw (10, -11.8) circle(.15);
\node[font=\fontsize{20}{6}\selectfont] at (9, -9.5) {$w_2$};
\end{tikzpicture}
\end{center}
\caption{An illustration of Operation~\ref{op02} where the dashed (thick, respectively) edges are in (not in, respectively) $\mc{S}$.
Operation~\ref{op02} converts two $2$-stars and two $3$-stars into a $4$-star and two $3$-stars.
\label{fig02}}
\end{figure}
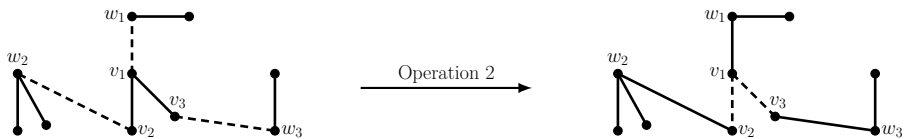

The second operation first finds a star $S = v_1$-$v_2\ldots v_\ell$ with order $\ell \in \{ 2, 3, 4 \}$ in $\mc{S}$
and $\ell$ separate critical vertices $\{ w_1, \ldots, w_\ell \}$ (see Definition~\ref{def01}) where no vertex $w_j$ appears in $S$.
If each vertex $v_j \in V(S)$ is adjacent to the critical vertex $w_j$, respectively,
then the operation replaces all edges of $S$ by the edges $\{ v_j, w_j \}_{j=1}^\ell$.
Figure~\ref{fig02} shows a representative case of Operation~\ref{op02} where in the left part, 
the central $3$-star is $S$ and each vertex $v_j$ is adjacent to the critical vertex $w_j$ for $j = 1, 2, 3$.
One sees that Operation~\ref{op02} is applicable, which revises two $2$-stars and two $3$-stars 
and; extracts a $4$-star and two $3$-stars.
We conclude Operation~\ref{op02} as follows.

\begin{operation}
\label{op02}
Suppose $S = v_1$-$v_2 \ldots v_\ell$ is a star with order $\ell \in \{ 2, 3, 4 \}$ in $\mc{S}$ and 
there exist $\ell$ separate critical vertices $\{ w_1, \ldots, w_\ell \}$
such that $w_j \notin V(S)$ for every $j = 1, \ldots, \ell$.
If $\{ v_j, w_j \}$ is an edge in $G$ for every $j$,
then the operation is applicable and replaces the edges of $S$ by the edges $\{ v_j, w_j \}_{j=1}^\ell$.
\end{operation}
Assume $w_j$ is in a tiny star $W_j$ and by Definition~\ref{def01}, $W_j$ is a $2$-star or $3$-star.
If $W_j$ is a $2$-star, then after Operation~\ref{op02}, $W_j$ becomes a $3$-star centered at $w_j$ (see $w_1$ in Figure~\ref{fig02}).
Otherwise, $W_j$ is a $3$-star centered at $w_j$ and it becomes a $4$-star after Operation~\ref{op02} (see $w_2$ in Figure~\ref{fig02}).
Therefore, we conclude that $\mc{S}$ remains a $k^-$-star partition after Operation~\ref{op02}.

We next discuss the running time of Operation~\ref{op02}.
Notice that exactly one star $S$ and $\ell \le 4$ separate critical vertices are considered by Operation~\ref{op02}.
Therefore, checking whether Operation~\ref{op02} is applicable can be done in $O(n^5)$ time.
The above analysis can be refined so that the running time is improved to $O(n^2)$.

To present the formal proof of $O(n^2)$, we fix a star $S = v_1$-$v_2 \ldots v_\ell$ with order $\ell \in \{ 2, 3, 4 \}$ for discussion.
Note that each vertex $w_j$ in Operation~\ref{op02} must be in a tiny star.
Therefore, we define the following tiny star types.

\begin{definition}
\label{def03}
(Tiny star types)
Suppose $S = v_1$-$v_2\ldots v_\ell$ is a star with order $\ell \in \{ 2, 3, 4 \}$.
Given a $2$-star $W \ne S$ in $\mc{S}$ where $V(W) = \{ w_1, w_2 \}$,
the {\em star type} of $W$ is a $(2\ell)$-tuple $(x_{11}, \ldots, x_{1\ell}, x_{21}, \ldots, x_{2\ell})$
where $x_{ij} \in \{ 0, 1 \}$ and $x_{ij} = 1$ if and only if $\{ w_i, v_j \}$ is an edge in $G$.

Similarly, given a $3$-star $W \ne S$ in $\mc{S}$ centered at $w$, the {\em star type} of $W$ is a $\ell$-tuple $(x_1, \ldots, x_\ell)$ where $x_j \in \{ 0, 1 \}$ and $x_j = 1$ if and only if $\{ w, v_j \}$ is an edge in $G$.
\end{definition}

By Definition~\ref{def03}, in the left part of Figure~\ref{fig02}, the star type of the top $2$-star is $(1, 0, 0, 0, 0, 0)$
and the leftmost $3$-star is $(0, 1, 0)$.
One sees that the tiny star type describes the edges connecting the (possible) center of a tiny star with each vertex of $S$.
Moreover, the tiny star type for each fixed tiny star can be determined in $O(1)$ time
and therefore, determining the tiny star types of every tiny star in $\mc{T} \setminus \{ S \}$ can be done in $O(n)$ time.

\begin{lemma}
\label{lemma02}
Given a tiny star $S \in \mc{S}$ with order $\ell$, there are at most $4^\ell + 2^\ell$ distinct tiny star types.
\end{lemma}
\begin{proof}
Clearly, the number of distinct star types of $2$-stars and $3$-stars is exactly $2^{2\ell}$ and $2^\ell$, respectively.
Then the proof is finished.
\end{proof}

Note that $\ell \in \{ 2, 3, 4 \}$ and thus the number of distinct tiny star types is bounded by a constant.
By Operation~\ref{op02}, every vertex $w_j$ must appear in a tiny star with a certain type.
One sees that once the tiny star type is fixed, then we can arbitrarily choose one star with the same type in $\mc{S} \setminus \{ S \}$.
Since each tiny star has at most three vertices, the vertex $w_j$ has at most $3(4^\ell + 2^\ell)$ choices.

\begin{lemma}
\label{lemma03}
Determining whether or not Operation~\ref{op02} is applicable can be done in $O(n^2)$ time 
and if Operation~\ref{op02} is applicable, then $q(\mc{S})$ is decreased by at least $4$.
\end{lemma}
\begin{proof}
We first prove the time complexity of Operation~\ref{op02}.
Operation~\ref{op02} first chooses a star $S$ in $\mc{S}$ with order $\ell \in \{ 2, 3, 4 \}$, which has at most $n$ choices.
Determining the types of all tiny stars can be done in $O(n)$ time.
Each $w_j$ may have at most $3(4^\ell + 2^\ell)$ choices
and thus finding the $\ell$ separate critical vertices $w_1, \ldots, w_\ell$ is in $3^\ell(4^\ell + 2^\ell)^\ell = O(1)$ time by $\ell \le 4$.

We next prove the change of $q(\mc{S})$.
For each vertex $w_j$, if it appears in a $2$-star ($3$-star, respectively), then after Operation~\ref{op02} is applied,
it appears in a $3$-star ($4$-star, respectively).
Therefore, for each $w_j$, the quantity $q(\mc{S})$ is reduced by at least $1$.
Note that $S$ disappears after Operation~\ref{op02}, which may be a $2$-, $3$- or $4$-star.
It is not hard to verify that $q(\mc{S})$ is decreased by at least $4$.
The lemma is proved.
\end{proof}

\begin{figure}[thb]
\begin{center}
\begin{tikzpicture}[scale=0.4,transform shape]

\draw [thick, line width = 1pt] (-8, -10) -- (-8, -12);
\draw [thick, line width = 1pt] (-8, -10) -- (-6.5, -11.5);
\filldraw (-8, -10) circle(.15);
\filldraw (-8, -12) circle(.15);
\filldraw (-6.5, -11.5) circle(.15);
\node[font=\fontsize{20}{6}\selectfont] at (-8.5, -10) {$v_1$};

\draw [densely dashed, line width = 1pt] (-8, -10) -- (-3, -10);
\draw [densely dashed, line width = 1pt] (-8, -10) -- (-3, -12);
\node[font=\fontsize{20}{6}\selectfont] at (-2.4, -10) {$w_1$};
\node[font=\fontsize{20}{6}\selectfont] at (-2.4, -12) {$w_2$};

\draw [thick, line width = 1pt] (-3, -10) -- (-3, -12);
\filldraw (-3, -10) circle(.15);
\filldraw (-3,-12) circle(.15);

\draw [-latex, thick] (0, -11) to (6, -11);
\node[font=\fontsize{20}{6}\selectfont] at (3, -10.5) {Operation~3};

\draw [thick, line width = 1pt] (9, -10) -- (9, -12);
\draw [thick, line width = 1pt] (9, -10) -- (10.5, -11.5);
\filldraw (9, -10) circle(.15);
\filldraw (9, -12) circle(.15);
\filldraw (10.5, -11.5) circle(.15);
\node[font=\fontsize{20}{6}\selectfont] at (8.5, -10) {$v_1$};

\draw [thick, line width = 1pt] (9, -10) -- (14, -10);
\draw [thick, line width = 1pt] (9, -10) -- (14, -12);
\node[font=\fontsize{20}{6}\selectfont] at (14.6, -10) {$w_1$};
\node[font=\fontsize{20}{6}\selectfont] at (14.6, -12) {$w_2$};

\draw [densely dashed, line width = 1pt] (14, -10) -- (14, -12);
\filldraw (14, -10) circle(.15);
\filldraw (14,-12) circle(.15);


\draw [thick, line width = 1pt] (-8, -14) -- (-8, -16);
\draw [thick, line width = 1pt] (-8, -14) -- (-6.5, -15.5);
\filldraw (-8, -14) circle(.15);
\filldraw (-8, -16) circle(.15);
\filldraw (-6.5, -15.5) circle(.15);
\node[font=\fontsize{20}{6}\selectfont] at (-8.5, -14) {$v_1$};
\node[font=\fontsize{20}{6}\selectfont] at (-8, -16.6) {$v_2$};

\draw [densely dashed, line width = 1pt] (-8, -14) -- (-3, -14);
\draw [densely dashed, line width = 1pt] (-8, -14) -- (-3, -16);
\node[font=\fontsize{20}{6}\selectfont] at (-2.4, -14) {$w_1$};
\node[font=\fontsize{20}{6}\selectfont] at (-2.4, -16) {$w_2$};

\draw [thick, line width = 1pt] (-3, -14) -- (-3, -16);
\filldraw (-3, -14) circle(.15);
\filldraw (-3,-16) circle(.15);

\draw [thick, line width = 1pt] (-11, -14) -- (-11, -16);
\filldraw (-11, -14) circle(.15);
\filldraw (-11,-16) circle(.15);
\draw [densely dashed, line width = 1pt] (-11, -16) -- (-8, -16);
\node[font=\fontsize{20}{6}\selectfont] at (-11, -16.6) {$w_3$};

\draw [-latex, thick] (0, -15) to (6, -15);
\node[font=\fontsize{20}{6}\selectfont] at (3, -14.5) {Operation~3};
\node[font=\fontsize{20}{6}\selectfont] at (3, -15.5) {$k=4$};

\draw [densely dashed, line width = 1pt] (12, -14) -- (12, -16);
\draw [thick, line width = 1pt] (12, -14) -- (13.5, -15.5);
\filldraw (12, -14) circle(.15);
\filldraw (12, -16) circle(.15);
\filldraw (13.5, -15.5) circle(.15);
\node[font=\fontsize{20}{6}\selectfont] at (11.5, -14) {$v_1$};
\node[font=\fontsize{20}{6}\selectfont] at (12, -16.6) {$v_2$};

\draw [thick, line width = 1pt] (12, -14) -- (17, -14);
\draw [thick, line width = 1pt] (12, -14) -- (17, -16);
\node[font=\fontsize{20}{6}\selectfont] at (17.6, -14) {$w_1$};
\node[font=\fontsize{20}{6}\selectfont] at (17.6, -16) {$w_2$};

\draw [densely dashed, line width = 1pt] (17, -14) -- (17, -16);
\filldraw (17, -14) circle(.15);
\filldraw (17,-16) circle(.15);

\draw [thick, line width = 1pt] (9, -14) -- (9, -16);
\filldraw (9, -14) circle(.15);
\filldraw (9,-16) circle(.15);
\draw [thick, line width = 1pt] (9, -16) -- (12, -16);
\node[font=\fontsize{20}{6}\selectfont] at (9, -16.6) {$w_3$};

\end{tikzpicture}
\end{center}
\caption{An illustration of Operation~\ref{op03} where the dashed (thick, respectively) edges are in (not in, respectively) $\mc{S}$.
The first line shows that Operation~\ref{op03} forms a $5$-star by merging a $3$-star and a $2$-star.
The second line shows that Operation~\ref{op03} revises a $3$-star and two $2$-stars and; forms a $4$-star and a $3$-star.
\label{fig03}}
\end{figure}
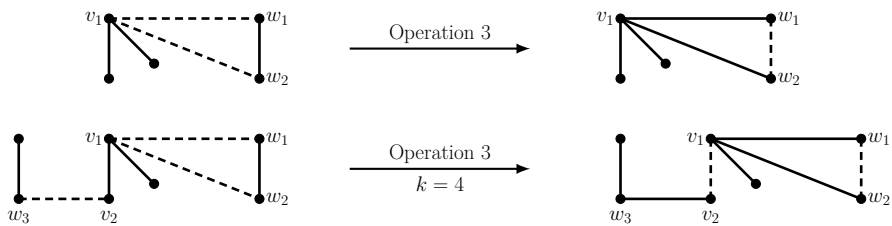

The third operation chooses a tiny star $S = v_1$-$v_2\ldots v_\ell$ in $\mc{S}$ and examines whether or not
there is a $2$-star $W = w_1$-$w_2$ in $\mc{S}$ such that $w_1, w_2$ are both adjacent to $v_1$.
If so, the operation replaces the edge $\{ w_1, w_2 \}$ by $\{ w_1, v_1 \}$ and $\{ w_2, v_2 \}$ unless $k = 4$ and $S$ is a $3$-star.
The first line of Figure~\ref{fig03} presents an illustration and we see that Operation~\ref{op03} forms a $5$-star by merging a $3$-star and a $2$-star.

If $k=4$ and $S$ is a $3$-star, then the operation further checks whether or not there is another critical vertex $w_3 \notin V(S) \cup V(W)$ such that $w_3$ is adjacent to $v_j$ for some $j \in \{ 2, 3 \}$.
If so, the operation replaces the edges $\{ w_1, w_2 \}$, $\{ v_1, v_j \}$ by $\{ w_1, v_1 \}$, $\{ w_2, v_1 \}$ and $\{ w_3, v_j \}$.
The second line of Figure~\ref{fig03} presents an illustration and we see that Operation~\ref{op03} forms a $4$-star and $3$-star from a $3$-star and two $2$-stars in $\mc{S}$.
Note that if $w_3$ is in $2$-star ($3$-star, respectively), then after Operation~\ref{op03}, $w_3$ appears in a $3$-star ($4$-star, respectively).
Therefore, $\mc{S}$ remains a $k^-$-star partition after Operation~\ref{op03}.

\begin{operation}
\label{op03}
Suppose that there is a tiny star $S=v_1$-$v_2 \ldots v_\ell$ and  a $2$-star $W = w_1$-$w_2$ in $\mc{S}$ 
such that $w_1, w_2 \in V(S)$ are both adjacent to $v_1$.
If $k \ge 5$ or $\ell = 2$, then the operation replaces the edge $\{ w_1, w_2 \}$ by $\{ w_1, v_1 \}$ and $\{ w_2, v_1 \}$.
Otherwise, the operation examines whether or not there is another critical vertex $w_3 \notin V(S) \cup V(W)$ such that $w_3$ is adjacent to $v_j$ for some $j \in \{ 2, 3 \}$.
If so, it replaces the edges $\{ w_1, w_2 \}$, $\{ v_1, v_j \}$ by $\{ w_1, v_1 \}$, $\{ w_2, v_1 \}$ and $\{ w_3, v_j \}$.
\end{operation}

\begin{lemma}
\label{lemma04}
Determining whether or not Operation~\ref{op03} is applicable can be done in $O(n^2)$ time 
and if Operation~\ref{op03} is applicable, then $q(\mc{S})$ is decreased by at least $3$.
\end{lemma}
\begin{proof}
The choices of $S$ is at most $n$ and similarly to Lemma~\ref{lemma03}, it takes $O(n)$ time to determine the tiny star types and whether or not Operation~\ref{op03} is applicable or not.

If Operation~\ref{op03} is applicable, $|\mc{S}_2|$ is reduced by at least one and $|\mc{S}_3|$ is at least non-increasing.
Therefore, the lemma is proved.
\end{proof}

Our algorithm starts with a $k^-$-star partition $\mc{S}$ with the least $1$-stars
and repeatedly perform Operations~\ref{op01}-\ref{op03} to update $\mc{S}$ until none of them is applicable.
We present the pseudo-code of our algorithm as follows. 

\begin{algorithm}
\caption{Algorithm {\sc Approx1} for $k^-$-star partition}
\label{Approx1}
\begin{algorithmic} 
\State Input: An undirected graph $G = (V, E)$;

\State Compute a $k^-$-star partition $\mc{S}$ with the least $1$-stars.

\While{One of Operations~\ref{op01}-\ref{op03} is applicable}
         \State Apply the operation to update $\mc{S}$.
\EndWhile

\State Return the $k^-$-star partition $\mc{S}$.
\end{algorithmic}
\end{algorithm}

\begin{theorem}
The algorithm {\sc Approx1} outputs a $k^-$-star partition in $O(n^3)$ time.
\end{theorem}
\begin{proof}
Clearly, {\sc Approx1} outputs a $k^-$-star partition.
By Lemma~\ref{lemma01}-\ref{lemma03} and $m \le n^2$, it takes at most $O(n^2)$ time to check whether one of Operations~\ref{op01}-\ref{op03} is applicable.
Since $0 \le q(\mc{S}) \le 4n$, by Lemma~\ref{lemma01}-\ref{lemma03} again, the while-loop in {\sc Approx1} can be executed at most $O(n)$ times.
In conclusion, {\sc Approx1} terminates in $O(n^3)$ time.
\end{proof}

\section{Approximation ratio}

In this section, we show the approximation ratio of {\sc Approx1}.
Note that the initial $k^-$-star partition $\mc{S}$ has the minimum $1$-stars and none of Operations~\ref{op01}-\ref{op03} generates a new $1$-star.
Therefore, the final returned $\mc{S}$ still has the minimum $1$-stars.

Recall that we fix an optimal $k^-$-star partition $\mc{Q}$ for discussion and let $\mc{Q}_j \subseteq \mc{Q}$ be the set of $j$-stars in $\mc{Q}$, $j = 1, \ldots, k$.
Therefore, we have the following equation.
\begin{equation}
\label{eq01}
|\mc{S}_1| \le |\mc{Q}_1|.
\end{equation}

We next present an amortization scheme where we assign a certain amount of tokens to each vertex based on the optimal solution $\mc{Q}$
so that we can estimate the number of tokens received by each star in $\mc{S}$.
For a fixed number $k \ge 4$, we set up a constant $\alpha(k)$, which is in the interval $(\frac 1k, \frac 1{k-1})$:
\begin{equation}
\label{eq02}
\alpha(k) = \frac {2k-3}{2k^2-4k+1} \in \left( \frac 1k, \frac 1{k-1} \right).
\end{equation}
We simplify $\alpha(k)$ to $\alpha$ since the value of $k$ is clear in the paper.

\begin{lemma}
\label{lemma05}
For a fixed $k \ge 4$ and its corresponding $\alpha$, we have

(i) $1-(k-1)\alpha = \frac {1-\alpha}{2(k-1)}$.

(ii) $1-(k-1)\alpha < \frac {1-\alpha}{k-1} < \frac 1k < \alpha < \frac {3(1-\alpha)}{2(k-1)}$.
\end{lemma}
\begin{proof}
(i) The equation is directly from Eq.(\ref{eq02}).

(ii) By Eq.(\ref{eq02}), we have $\alpha > \frac 1k$ and thus $1-(k-1)\alpha < \frac {1-\alpha}{k-1} < \frac 1k < \alpha$.
Since $k \ge 4$, one sees that $\frac 1{k-1} \le \frac 3{2k+1}$
and thus the inequation $\alpha < \frac {3(1-\alpha)}{2(k-1)}$ follows from $\alpha < \frac 3{2k+1}$.
\end{proof}

We are ready to present our amortization scheme.

\begin{definition}
\label{def04}
(Amortization scheme)
We assign $\alpha$ token to the vertex of each $1$-star in $\mc{Q}_1$.

We next consider a $j$-star $Q = v_1$-$v_2 \ldots v_j$ in $\mc{Q}_j$ with $2 \le j \le k$.
We distinguish the following three cases:
\begin{itemize}
\item[1.] $v_1$ is critical (see Definition~\ref{def01}).
We assign $\alpha$ token to $v_1$ and $\frac {1-\alpha}{j-1}$ token to each of $\{ v_2, \ldots, v_j \}$, respectively.

\item[2.] $v_1$ is not critical and at least one of $\{ v_2, \ldots, v_j \}$ is critical.
We assign $\alpha$ token to each of $\{ v_2, \ldots, v_j \}$ and $1-(j-1)\alpha$ token to $v_1$, respectively.

\item[3.] None of $\{ v_1, \ldots, v_j \}$ is critical.
We assign $\frac 1j$ token to each of $\{ v_1, \ldots, v_j \}$, respectively.
\end{itemize}

\end{definition}

\begin{lemma}
\label{lemma06}
A vertex receives at least $\frac {1-\alpha}{k-1}$ token if it is a satellite of $\mc{Q}$ or it is critical.
Moreover, each vertex receives at least $1-(k-1)\alpha$ token.
\end{lemma}
\begin{proof}
We consider a vertex $v$ and we assume $v$ is in a $j$-star $Q$ in $\mc{Q}$.
If $j = 1$, then $v$ receives $\alpha$ token and by Lemma~\ref{lemma05}(ii), the lemma is proved.
Therefore, we can assume $Q$ is a $2^+$-star.

If $v$ is a satellite of $Q$, then by Definition~\ref{def04}, $v$ receives at least $\min \{ \frac {1-\alpha}{j-1}, \alpha, \frac 1j \}$ token.
The lemma is proved by $j \le k$ and Lemma~\ref{lemma05}(ii).
Similarly, we can prove the remaining part of the lemma.
\end{proof}

We next analyze how many tokens a star in $\mc{S}$ can receive based on the orders.

\begin{lemma}
\label{lemma07}
A $4^+$-star in $\mc{S}$ receives at least $\frac {5(1-\alpha)}{2(k-1)}$ token.
\end{lemma}
\begin{proof}
Consider a $4^+$-star $S \in \mc{S}$.
By Lemma~\ref{lemma06}, if the order of $S$ is at least $5$, then it receives at least $5(1-(k-1)\alpha)$ token.
By Lemma~\ref{lemma05}(i), the lemma is proved.

We next assume $S$, denoted as $v_1$-$v_2v_3v_4$, is a $4$-star in $\mc{S}$
and the vertex $v_j$ is in a star $Q_j \in \mc{Q}$ for every $j \in \{ 1, 2, 3, 4 \}$.
We discuss the following three cases.

Case 1: At least one of $Q_j$ is a $1$-star.
Then $v_j$ receives $\alpha$ token.
By Lemma~\ref{lemma06}, $S$ receives at least $\alpha + 3(1-(k-1)\alpha)$ token
and by Lemma~\ref{lemma05}(ii), we are done. 

We next assume every star $Q_j$ is a $2^+$-star.

Case 2: At least one vertex $v_j$ is a satellite in $\mc{Q}$.
By Lemma~\ref{lemma06}, at least $\frac {1-\alpha}{k-1}$ token is assigned to $v_j$
and by Lemma~\ref{lemma06} again, $S$ receives at least $\frac {1-\alpha}{k-1} + 3(1-(k-1)\alpha)$ token.
By Lemma~\ref{lemma05}(i), the lemma is proved. 

Case 3: Every vertex of $S$ is a center in $\mc{Q}$. 
Note that every vertex $v_j$ is not critical since $S$ is a $4$-star.

Case 3.1: At least one of $Q_j$ has no critical vertex.
In this case, by Definition~\ref{def04}, $v_j$ can get at least $\frac 1k$ token, which happens when $Q_j$ is exactly a $k$-star.
By Lemma~\ref{lemma05}(ii), we have $\frac 1k > \frac {1-\alpha}{k-1}$ and thus similarly to Case 2, we are done.

Case 3.2:  Every $Q_j$ has a critical vertex.
Since $v_j$ is not critical, we assume $w_j \ne v_j$ is a critical vertex in $Q_j$.
The star $Q_j$ is centered at $v_j$ and thus $\{ v_j, w_j \}$ is an edge of $G$.
One sees that $w_j \notin V(S)$ since no vertex in $S$ is critical.

We next prove that the found $w_1, w_2, w_3, w_4$ are separate (see Definition~\ref{def01}).
Clearly $w_2$ is not in a $2$-star, otherwise, Operation~\ref{op01} is applicable and so are $w_3, w_4$.
Therefore, by Definition~\ref{def01}, each of $w_2, w_3, w_4$ is the center of some $3$-star in $\mc{S}$, respectively.
One sees that $w_1$ is either in a $2$-star or the center of $3$-star in $\mc{S}$.
We conclude that $w_1, w_2, w_3, w_4$ are separate.
It follows that Operation~\ref{op02} is applicable, a contradiction.
\end{proof}

Before we discuss the tokens assigned to $3$-stars or $2$-stars in $\mc{S}$, 
we distinguish two special pairs of $2$- or $3$-stars in $\mc{S}$.

\begin{definition}
\label{def05}
A $3$-star $S = v_1$-$v_2v_3$ in $\mc{S}$ is defined as {\em special} if 
there exists a $2$-star $W = w_1$-$w_2$ in $\mc{S}$ such that the following two conditions are satisfied.
\begin{itemize}
\item[C1.] $v_1$ is a satellite of $\mc{Q}$ and $v_2, v_3$ are two centers of $\mc{Q}$;

\item[C2.] there exists two vertices $v_i, v_j \in V(S)$ such that $\{ v_i, w_1 \}$ and $\{ v_j, w_2 \}$ are two edges in $\mc{Q}$.
\end{itemize}

A $2$-star $S = v_1$-$v_2$ is defined as {\em special} if $v_1$ and $v_2$ are both satellites in $\mc{Q}$
and {\em one} of the following conditions is satisfied.
\begin{itemize}
\item[C3.] $k=4$ and there exists a $3$-star centered at $w$ such that $\{ v_1, w \}$ and $\{ v_2, w \}$ are two edges in $\mc{Q}$.
That is, $w$ is a center of both $\mc{S}$ and $\mc{Q}$.

\item[C4.] There exists a $2$-star $W = w_1$-$w_2$ 
such that $\{ v_1, w_1 \}$ and $\{ v_2, w_2 \}$ are two edges in $\mc{Q}$.
\end{itemize}

In the above two cases, we say the $2$-star $W$ in the conditions C1, C2, C4 or the $3$-star in C3 is {\em associated} with $S$.
A $2$- or $3$-star is {\em regular} if it is not a special.
\end{definition}

\begin{figure}[thb]
\begin{center}
\begin{tikzpicture}[scale=0.5,transform shape]

\draw [thick, line width = 1pt] (-8, -12) -- (-5.5, -11);
\draw [thick, line width = 1pt] (-8, -10) -- (-5.5, -11);
\filldraw (-8, -10) circle(.15);
\filldraw (-8, -12) circle(.15);
\filldraw (-5.5, -11) circle(.15);
\node[font=\fontsize{20}{6}\selectfont] at (-5.5, -11.6) {$v_1$};
\node[font=\fontsize{20}{6}\selectfont] at (-8, -10.6) {$v_2$};
\node[font=\fontsize{20}{6}\selectfont] at (-8, -12.6) {$v_3$};

\draw [densely dashed, line width = 1pt] (-8, -10) -- (-3, -10);
\draw [densely dashed, line width = 1pt] (-5.5, -11) -- (-3, -12);

\draw [thick, line width = 1pt] (-3, -10) -- (-3, -12);
\filldraw (-3, -10) circle(.15);
\filldraw (-3,-12) circle(.15);
\node[font=\fontsize{20}{6}\selectfont] at (-2.4, -10) {$w_2$};
\node[font=\fontsize{20}{6}\selectfont] at (-2.4, -12) {$w_1$};

\draw [densely dashed, line width = 1pt] (-10, -10) -- (-8, -10);
\draw [densely dashed, line width = 1pt] (-10, -12) -- (-8, -12);

\draw [thick, line width = 1pt] (-10, -10) -- (-10, -12);
\filldraw (-10, -10) circle(.15);
\filldraw (-10,-12) circle(.15);
\node[font=\fontsize{20}{6}\selectfont] at (-10.6, -10) {$w'_1$};
\node[font=\fontsize{20}{6}\selectfont] at (-10.6, -12) {$w'_2$};

\draw [thick, line width = 1pt] (5, -10) -- (5, -12);
\filldraw (5, -10) circle(.15);
\filldraw (5, -12) circle(.15);
\node[font=\fontsize{20}{6}\selectfont] at (4.4, -10) {$v_1$};
\node[font=\fontsize{20}{6}\selectfont] at (4.4, -12) {$v_2$};

\draw [densely dashed, line width = 1pt] (5, -10) -- (8, -10);
\draw [densely dashed, line width = 1pt] (5, -12) -- (8, -12);

\draw [thick, line width = 1pt] (8, -10) -- (8, -12);
\filldraw (8, -10) circle(.15);
\filldraw (8,-12) circle(.15);
\node[font=\fontsize{20}{6}\selectfont] at (8.6, -10) {$w_1$};
\node[font=\fontsize{20}{6}\selectfont] at (8.6, -12) {$w_2$};

\end{tikzpicture}
\end{center}
\caption{An illustration of a special $3$-star and $2$-star and; their associated $2$-stars,
where the dashed (thick, respectively) edges are in $\mc{S}$ (in $\mc{Q}$, respectively).
In the left part, the central $3$-star $v_1$-$v_2v_3$ is special and it associates with $w_1$-$w_2$ and $w'_1$-$w'_2$.
In the right part, the leftmost $2$-star $v_1$-$v_2$ is special and associates with $w_1$-$w_2$
and vice versa.
\label{fig04}}
\end{figure}
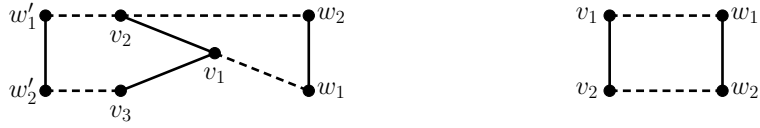

Figure~\ref{fig04} shows a possible special $3$-star and $2$-star as well as their associated $2$-stars where the thick and dashed edges are in $\mc{S}$ and $\mc{Q}$, respectively.
For example, in the left part, there is a $3$-star $v_2$-$w'_1w_2$, two $2$-stars $v_3$-$w'_2$ and $w_1$-$v_1$ in the optimal solution.
Therefore, one sees that the central $3$-star is special, which associates the two $2$-stars $w'_1$-$w'_2$ and $w_1$-$w_2$.

Note that if a $3$-star $W$ is associated with a special $2$-star, then by C3, the center of $W$ is a center of $\mc{Q}$ and thus $W$ is regular.
Similarly, we can verify that a $2$-star $W$ associated with a special star is regular.
In Figure~\ref{fig04}, we see that a special $3$-star may associate with multiple $2$-stars.
However, the following lemma shows that a $2$-star or $3$-star in $\mc{S}$ can be associated with at most a special star.

\begin{lemma}
\label{lemma08}
A $2$-star or a $3$-star in $\mc{S}$ can be associated with exactly one special star.
\end{lemma}
\begin{proof}
We first assume $W= w_1$-$w_2w_3$ is a $3$-star in $\mc{S}$, which is associated with two distinct special stars $S, S' \in \mc{S}$.
By the condition C3 in Definition~\ref{def05}, $k = 4$ and $S, S'$ are both $2$-stars.
Therefore, $w_1$ is a center of a $5^+$-star in $\mc{Q}$ since $\{ w_1, v \}$ is an edge in $\mc{Q}$ for every vertex $v \in V(S) \cup V(S')$.
This contradicts that $\mc{Q}$ is a $4^-$-star partition and we are done.

We next assume $W= w_1$-$w_2$ is a $2$-star in $\mc{S}$, which is associated with two distinct special stars $S, S' \in \mc{S}$.
In this case, each of $w_1, w_2$ is incident to two edges in $\mc{Q}$.
In other words, $w_1, w_2$ are two centers in $\mc{Q}$.
If one of $S, S'$ is a $3$-star, then by C1 and C2 in Definition~\ref{def05}, at least one of $w_1, w_2$, is a satellite in $\mc{Q}$, a contradiction.
Therefore, we can assume that $S, S'$ are two $2$-stars in $\mc{S}$, denoted as $S = v_1$-$v_2$ and $S' = v'_1$-$v'_2$, respectively.
Note that the condition C4 is satisfied
and without loss of generality, we assume $\{ w_1, v_1 \}, \{ w_1, v'_1 \}, \{ w_2, v_2 \}$ and $\{ w_2, v'_2 \}$ are the edges in $\mc{Q}$.
One sees that Operation~\ref{op02} is applicable, a contradiction again.

The lemma is proved.
\end{proof}

For a special star, if multiple stars are associated with it, then we arbitrarily choose one.
This way, by Lemma~\ref{lemma08}, we can assume every special star associates with a one-to-one star in $\mc{S}$. 
In the following lemmas, we show that every $2^+$-star in $\mc{S}$ receives at least $\alpha+\frac {1-\alpha}{k-1} = \frac {4k-7}{2k^2-4k+1}$ token on average,
which leads to the final approximation ratio.

\begin{lemma}
\label{lemma09}
A special $2$-star and its one-to-one associated star can receive at least $2\alpha + \frac {2-2\alpha}{k-1}$ token.
\end{lemma}
\begin{proof}
We first consider a special $2$-star $S = v_1$-$v_2$ and its associated star $W = w_1$-$w_2 \ldots w_\ell$ with $\ell \in \{ 2, 3 \}$.
Note that $v_1, v_2$ are critical and thus by Lemma~\ref{lemma06}, they receive at least $\frac {2-2\alpha}{k-1}$ token in total.
We discuss the following two cases based on the value of $\ell$.

Case 1: $W$ is a $3$-star, i.e., $\ell = 3$.
Since $w_1$ is critical and a center in $\mc{Q}$, by Definition~\ref{def04}, $w_1$ receives $\alpha$ token.
We assume $w_j$ is in a star $Q_j \in \mc{Q}$, $j = 2, 3$.

Case 1.1: $w_2$ or $w_3$ is a satellite of $\mc{Q}$.
By Lemma~\ref{lemma06}, $w_2$ and $w_3$ receives at least $\frac {1-\alpha}{k-1} + 1-(k-1)\alpha$ token.
Therefore, by Lemma~\ref{lemma05}(i) and (ii), the token received by the vertices of $W$ and $S$ is at least
\[
\alpha + \frac {3-3\alpha}{k-1} + 1-(k-1)\alpha = \alpha + \frac {7-7\alpha}{2k-2} 
> 2\alpha + \frac {2-2\alpha}{k-1},
\]
which completes the proof.

Case 1.2:  $w_2$ and $w_3$ are two centers of $\mc{Q}$ and; $Q_2$ and $Q_3$ do not have a critical vertex.
In this case, by Definition~\ref{def04}, $w_2, w_3$ receives at least $\frac 2k$ token, which is greater than $\frac {1-\alpha}{k-1} + 1-(k-1)\alpha$ by Lemma~\ref{lemma05}(ii).
Similarly to Case 1.1, we are done.

Case 1.3: $w_2$ and $w_3$ are two centers of $\mc{Q}$ and; $Q_2$ or $Q_3$, say $Q_2$ has a critical vertex $v$.
Since $w_2$ is not critical,  we have $w_2 \ne v$ and $\{ w_2, v \}$ is an edge in $G$.
Clearly, $v$ is not in $V(S) \cup V(W)$ and thus Operation~\ref{op03} is applicable, a contradiction.

Case 2: $W$ is a $2$-star, i.e., $\ell = 2$.
By Definition~\ref{def01}, $v_1, v_2, w_1, w_2$ are critical and by Definition~\ref{def05}, $w_1, w_2$ are two centers in $\mc{Q}$.
By Definition~\ref{def04} and Lemma~\ref{lemma06}, $w_1, w_2$ receive $2\alpha$ token and $v_1, v_2$ receive at least $\frac {2-2\alpha}{k-1}$ token in total, respectively.
We are done.

Together with all cases, the proof is finished.
\end{proof}

\begin{lemma}
\label{lemma10}
A special $3$-star and its one-to-one associated star can receive at least $2\alpha + \frac {2-2\alpha}{k-1}$ token.
\end{lemma}
\begin{proof}
Note that the associated star must be a $2$-star.
Therefore, we can consider a special $3$-star $S = v_1$-$v_2v_3$ and its associated $2$-star $W = w_1$-$w_2$.
By Definition~\ref{def01}, $v_1$ is critical and by Lemma~\ref{lemma06}, $v_1$ receives at least $\frac {1-\alpha}{k-1}$ token
and $v_2, v_3$ can receive at least $2-2(k-1)\alpha$ token in total.
By Definition~\ref{def05}, we discuss the following three cases.

Case 1: $\{ v_2, w_1 \}$ and $\{ v_3, w_2 \}$ are two edges in $\mc{Q}$.
Note that each of $w_1, w_2$ receives $\alpha$ token since they are critical and satellites in $\mc{Q}$ and;
their corresponding centers in $\mc{Q}$, i.e., $v_2, v_3$ are not critical.
By Lemma~\ref{lemma05}(i), at least $\frac {1-\alpha}{k-1} + 2-2(k-1)\alpha + 2\alpha = 2\alpha + \frac {2-2\alpha}{k-1}$ 
token in total is assigned to $S$ and $W$.
The proof is completed.

Case 2: $\{ v_1, w_1 \}$ and $\{ v_2, w_2 \}$ are two edges in $\mc{Q}$.
Similarly to Case 1, we can verify that $w_2$ receives $\alpha$ token.
Note that in this case, $w_1$ is a center in $\mc{Q}$ and $w_1$ is a critical vertex.
Therefore, $w_1$ receives at least $\alpha$ token.
One sees that the vertices of $S$ and $W$ receive at least $\frac {1-\alpha}{k-1} + 2-2(k-1)\alpha + 2\alpha = 2\alpha + \frac {2-2\alpha}{k-1}$ token in total.

Case 3: $\{ v_1, w_1 \}$ and $\{ v_3, w_2 \}$ are two edges in $\mc{Q}$.
The case is symmetrical with Case 2.
Together with these cases, the proof is done.
\end{proof}

\begin{lemma}
\label{lemma11}
A regular $3$-star in $\mc{S}$ receives at least $\alpha + \frac {1-\alpha}{k-1}$ token.
\end{lemma}
\begin{proof}
Let $S=v_1$-$v_2v_3$ denote a regular $3$-star in $\mc{S}$ and thus $v_1$ is critical.
We assume $v_j$ is in a star $Q_j \in \mc{Q}$ for every $j = 1, 2, 3$ and discuss the following three cases.

Case 1: At least one of $Q_j$ is a $1$-star.
Then by Definition~\ref{def04}, $v_j$ receives $\alpha$ token.
By Lemma~\ref{lemma06} and Lemma~\ref{lemma05}(i), the vertices in $V(S) \setminus \{ v_j \}$ receive at least $2(1-(k-1)\alpha) = \frac {1-\alpha}{k-1}$ token, which proves the lemma.

We next assume every $Q_j$ is a $2^+$-star.
Let $w_1$ be the center of $Q_1$.

Case 2: $v_1 = w_1$ or; $v_1 \ne w_1$ and $w_1$ is not critical.
If $v_1 = w_1$, then since $v_1$ is critical, by Definition~\ref{def04}, $v_1$ receives $\alpha$ token.
In the other case, $v_1$ is a satellite of $Q_1$ centered at $w_1$.
Since $v_1$ is critical and $w_1$ is not critical, by Definition~\ref{def04} again, $v_1$ receives $\alpha$ token.
Therefore, we conclude that $v_1$ always receives $\alpha$ token
and thus similarly to Case 1, we are done.

Case 3:  $v_1 \ne w_1$ and $w_1$ is critical.
Therefore, $\{ v_1, w_1 \}$ is an edge of $\mc{Q}$.
Since $w_1$ is critical and none of $v_2, v_3$ is critical, $w_1 \notin V(S)$.
By Lemma~\ref{lemma06}, at least $\frac {1-\alpha}{k-1}$ token is assigned to $v_1$.

Case 3.1: One of $v_2, v_3$ is a satellite of $\mc{Q}$.
By Lemma~\ref{lemma06} and Lemma~\ref{lemma05}(i), the vertices of $S$ can receive at least $\frac {2-2\alpha}{k-1} + 1-(k-1)\alpha = \frac {5-5\alpha}{2k-2}$ token,
which is greater than $\alpha + \frac {1-\alpha}{k-1}$ by Lemma~\ref{lemma05}(ii).

Case 3.2: Each of $v_2, v_3$ is a center of $\mc{Q}$ and $Q_2, Q_3$ have no critical vertex.
Therefore, $v_2, v_3$ receive at least $\frac 2k$ token
and thus the vertices of $S$ can receive $\frac 2k + \frac {1-\alpha}{k-1}$ token.
The lemma is proved due to Eq.(\ref{eq02}) and $\frac 1{k-1} < \frac 2k$ by $k \ge 4$.

Case 3.3: Each of $v_2, v_3$ is a center of $\mc{Q}$ and each of $Q_2, Q_3$ has a critical vertex, denoted as $w_2, w_3$, respectively.
In this case, $\{ v_j, w_j \}$ is an edges of $\mc{Q}$, for every $j \in \{ 1, 2, 3 \}$.
One sees that $w_1, w_2, w_3$ are critical vertices but $v_2, v_3$ are not critical.
Recall that $w_1 \notin V(S)$ and $v_1$ is a satellite of $w_1$.
Thus, $w_2, w_3 \ne v_1$.
Therefore, each of $w_1, w_2, w_3$ is not in $V(S)$.

We next prove $w_1, w_2, w_3$ are separate.
If not, then by Definition~\ref{def01}, we can assume there exist $i, j \in \{ 1, 2, 3 \}$ such that $w_i$ and $w_j$ are in a $2$-star of $\mc{S}$.
One sees that if such $2$-star exists, then $S$ becomes a speical $3$-star, a contradiction.
Therefore $w_1, w_2, w_3$ are separate, which is impossible since Operation~\ref{op02} is applicable, a contradiction again.

Together with the three cases, we prove the lemma.
\end{proof}

\begin{lemma}
\label{lemma12}
A regular $2$-star in $\mc{S}$ receives at least $\alpha + \frac {1-\alpha}{k-1}$ token.
\end{lemma}
\begin{proof}
We consider a $2$-star $S = v_1$-$v_2$ in $\mc{S}$ and we discuss the following two cases.

Case 1: One of $v_1, v_2$ is a center of $\mc{Q}$.
Therefore, by Definition~\ref{def04} and Lemma~\ref{lemma06}, the number of tokens received by $v_1, v_2$ is at least $\alpha + \frac {1-\alpha}{k-1}$ and we are done.

Case 2: $v_1, v_2$ are two satellites of $\mc{Q}$.
We assume $v_j$ is in a star $Q_j \in \mc{Q}$ centered at $w_j$ for $j \in \{ 1, 2 \}$.
Note that $Q_1, Q_2$ are both $2^+$-stars since $v_1, v_2$ are their satellites.
Clearly, $w_1, w_2 \notin V(S)$.

Case 2.1: One of $w_1, w_2$, say $w_1$, is not critical.
Since $v_1$ is critical, by Definition~\ref{def04}, $v_1$ receives $\alpha$ token.
The lemma follows from Lemma~\ref{lemma06} and $v_2$ is critical.

Case 2.2: $w_1, w_2$ are both critical and $w_1 = w_2$.
Note that the case does not exist if $k \ge 5$ or $w_1$ is in a $2$-star in $\mc{S}$; otherwise, Operation~\ref{op03} is applicable.
Therefore, we can assume $k = 4$ and $w_1$ is the center of a $3$-star in $\mc{S}$.
Then $S$ is a special $2$-star by the condition C3, a contradiction again.

Case 2.3: $w_1, w_2$ are both critical and $w_1 \ne w_2$.
Since $S$ is regular, by the condition C4, $w_1, w_2$ cannot be in a same $2$-star in $\mc{S}$.
Therefore, $w_1, w_2$ are separate, which indicates that Operation~\ref{op02} is applicable again, a contradiction.

Together with these two cases, the lemma is proved.
\end{proof}

\begin{theorem}
\label{thm02}
Algorithm {\sc Approx1} is at most $\frac {2k^2-4k+1}{4k-7}$-approximation for every $k \ge 4$.
\end{theorem}
\begin{proof}
Note that by Definition~\ref{def04}, every $1$-star in $\mc{Q}$ has $\alpha$ token and every $2^+$-star in $\mc{Q}$ has exactly one token.
Therefore, the total number of tokens is exactly $\alpha |\mc{Q}_1| + |\mc{Q}| - |\mc{Q}_1| = |\mc{Q}| + (\alpha-1) |\mc{Q}_1|$.

On the other hand, by Lemmas~\ref{lemma07}-\ref{lemma12}, each $2^+$-star in $\mc{S}$ receives $\alpha + \frac {2-2\alpha}{k-1}$ token on average.
Therefore, we can conclude that the number of received tokens is at least $( \alpha + \frac {1-\alpha}{k-1} ) ( |\mc{S}| - |\mc{S}_1| )$.
It follows that
\begin{equation}
\label{eq03}
( \alpha + \frac {1-\alpha}{k-1} ) ( |\mc{S}| - |\mc{S}_1| ) 
\le |\mc{Q}| + (\alpha-1) |\mc{Q}_1|.
\end{equation}
Using Eqs.(\ref{eq01}), (\ref{eq03}) and $k \ge 4$,  we finally obtain 
\begin{eqnarray*}
( \alpha + \frac {1-\alpha}{k-1} ) |\mc{S}| 
& \le & |\mc{Q}| + (\alpha-1) |\mc{Q}_1| + (\alpha + \frac {1-\alpha}{k-1}) |\mc{S}_1| \\
& \le &  |\mc{Q}| + (2\alpha- 1 + \frac {1-\alpha}{k-1}) |\mc{Q}_1|  \\
& < &  |\mc{Q}| + (\frac 2{k-1} - 1 + \frac 1{k-1}) |\mc{Q}_1|  \\
& = & |\mc{Q}| + \frac {4-k}{k-1} |\mc{Q}_1|  \le |\mc{Q}|.
\end{eqnarray*}
The theorem is proved since $\alpha + \frac {1-\alpha}{k-1} = \frac {4k-7}{2k^2-4k+1}$.
\end{proof}

\section{Conclusions}

In this paper, we study the $k^-$-star partition problem in an undirected graph $G=(V, E)$
and present an improved $O(|V|^3)$-time $(\frac k2 - \frac {k-2}{8k-14})$-approximation for every $k \ge 4$ based on local search.
Our algorithm starts with a $k^-$-star partition with the minimum number of $1$-stars and 
converts it to a solution by repeatedly applying three local search operations to the current solution.
We distinguish the critical vertices from the solution and allow those vertices to receive more tokens in our amortized analysis.

One sees that when $k$ tends to infinity, the approximation ratio is still approaching $\frac k2-\frac 18$.
Therefore, for future research, one direction is to improve the approximation ratio especially for a large $k$
or to prove the lower bound on the problem.
Another direction is to extend the problem to a directed graph by considering directed stars.




\bibliography{output}

\end{document}